\documentclass[final,authoryear,5p,times,twocolumn]{elsarticle}

\usepackage{savesym}
\savesymbol{AND}

\usepackage{amsmath,amsfonts,amssymb}
\usepackage{graphicx}
\usepackage{multirow}
\usepackage{subfigure}
\usepackage{hyperref}
\usepackage{libertine}

\usepackage{algorithm}
\usepackage{algorithmic}
\usepackage{bm}
\usepackage{upgreek}

\graphicspath{{figures/}}

\DeclareMathOperator*{\argmin}{arg\,min}
\DeclareMathOperator*{\vecop}{vec}

\usepackage[usenames, dvipsnames]{xcolor}
\usepackage[normalem]{ulem}
\usepackage{todonotes}

\usepackage{color, colortbl}

\newcolumntype{C}[1]{>{\centering\let\newline\\\arraybackslash\hspace{0pt}}m{#1}}

\def\B#1{\mathbf{#1}}

\def\bfbeta{{\boldsymbol{\upbeta}}}
\def\bfomega{{\boldsymbol \upomega}}
\def\bfalpha{{\boldsymbol \upalpha}}

\def\slantfrac#1#2{\kern.1em^{#1}\kern-.1em/\kern-.1em_{#2}}

\def\RR{{\mathbb R}}

\let\oldparagraph\paragraph
\def\paragraph#1{\oldparagraph{\bf #1}}

\renewcommand\cite{\citep}

\begin{document}

\begin{frontmatter}

\title{Data-driven HRF estimation for encoding and decoding models}
% \title{Data-driven HRF estimates for encoding and decoding}
% \title{Don't use a canonical HRF for encoding and decoding}
% \title{Rank-1 models for HRF estimation with validation on encoding and decoding}

\author{Fabian Pedregosa\corref{cor0}\fnref{equal,INRIA,neurospin}}
\author{Michael Eickenberg\fnref{equal,INRIA,neurospin}}
\author{Philippe Ciuciu\fnref{INRIA,neurospin}}
\author{Bertrand Thirion\fnref{INRIA,neurospin}}
\author{Alexandre Gramfort\fnref{paristech,neurospin}}

% % use optional labels to link authors explicitly to addresses:
\fntext[equal]{Authors contributed equally}
\fntext[INRIA]{Parietal Team, INRIA Saclay-\^{I}le-de-France, Saclay, France}
\fntext[paristech]{Institut Mines-Telecom, Telecom ParisTech, CNRS LTCI, 37-39 Rue Dareau, 75014 Paris, France}
\fntext[neurospin]{NeuroSpin, CEA Saclay, Bat. 145, 91191 Gif-sur-Yvette Cedex, France}
\cortext[cor0]{Corresponding author. Email: fabian.pedregosa@inria.fr. Tel: +33 1 69 08 79 92}

%%%%%%%%%%%%%%%%%%%%%%%%%%%%%%%%%%%%%%%%%%%%%%%%%%%%%%%%%%%%%%%%%%%%%%%%%%%%%%%%

\begin{abstract} Despite the common usage of a canonical, data-independent,
hemodynamic response function (HRF), it is known that the shape of the HRF
varies across brain regions and subjects. This suggests that a data-driven
estimation of this function could lead to more statistical power when modeling
BOLD fMRI data. However, unconstrained estimation of the HRF can yield highly
unstable results when the number of free parameters is large. We develop a
method for the joint estimation of activation and HRF by means of a rank constraint, 
forcing the estimated HRF to be equal across events or experimental conditions,
 yet permitting it to differ across voxels. Model estimation leads to
an optimization problem that we propose to solve with an efficient
\mbox{quasi-Newton} method, exploiting fast gradient computations. This model,
called GLM with Rank-1 constraint (R1-GLM), can be extended to the setting of
GLM with separate designs which has been shown to improve decoding accuracy in brain activity decoding experiments.
We compare 10 different HRF modeling methods in terms of encoding and decoding
score on two different datasets. Our results show that the \mbox{R1-GLM} model
outperforms competing methods in both encoding and decoding
settings, positioning it as an attractive method both from the points of view
of accuracy and computational efficiency.
\end{abstract}


\end{frontmatter}

\section{Introduction}

The use of machine learning techniques to
predict the cognitive state of a subject from their functional MRI (fMRI)
data recorded during task performance has become a
popular analysis approach for neuroimaging
studies over the last decade~\cite{Cox2003, Haynes2006}.
It is now commonly referred to as \emph{brain reading} or \emph{decoding}.
In this setting, the BOLD signal is used to predict the
task or stimulus that the subject was performing.
Although it is possible to perform decoding directly on raw BOLD
signal~\cite{Mourao-Miranda2007, Miyawaki2008}, the common approach in 
fast event-related designs consists in extracting the activation coefficients (beta-maps) from
the BOLD signal to perform the decoding analysis on these estimates.
Similarly, in the voxel-based \emph{encoding} models~\cite{Kay2008, Naselaris2011}, the activation coefficients are extracted from the BOLD signal,
this time to learn a model to predict the BOLD response in
a given voxel, based on a given representation of the stimuli. In addition, a third approach, known as representational similarity analysis or RSA~\cite{kriegeskorte2008representational} takes as input the activation coefficients. In this case a comparison is made between the similarity observed in the activation coefficients, quantified by a correlation measure, and the
 similarity between the stimuli, quantified by a similarity measure defined from
the experimental setting.

These activation coefficients are computed by means of the General Linear Model
(GLM)~\cite{Friston1995}. While
this approach has been successfully used in a wide range of studies, it does
suffer from limitations~\cite{Poline2012}. For instance, the GLM commonly
relies on a \mbox{data-independent} \emph{canonical} form of the hemodynamic response function
(HRF) to estimate the activation coefficient. However it is
known~\cite{Handwerker2004,Badillo2013} that the shape of this response function
can vary substantially across subjects and brain regions. This suggests that an adaptive modeling of this
 response function should improve the accuracy of subsequent analysis.

To overcome the aforementioned limitation, Finite Impulse Response (FIR) models have been
proposed within the GLM framework~\cite{Dale1999,Glover1999}.
These models do not assume any particular shape for the HRF and amount to
estimating a large number of parameters in order to identify it. 
While the FIR-based modeling makes it possible to estimate the
activation coefficient and the HRF simultaneously, the increased flexibility
has a cost. The estimator is less robust and prone to overfitting, i.e. to generalize badly to unseen data. 
In general, FIR
models are most appropriate for studies focused on the characterization of the
shape of the hemodynamic response, and not for studies that are primarily
focused on detecting activation~\cite[Chapter~5]{Poldrack}.

Several strategies aiming at reducing the number of degrees of freedom of the
FIR model - and thus at limiting the risk of overfitting - have been proposed.
One possibility is to constrain the shape of the HRF to be a linear
combination of a small number of basis functions. A common choice of basis is 
formed by three elements consisting of a reference HRF as well as its time and dispersion
derivatives~\cite{friston1998nonlinear}, although it is also possible to compute a
basis set that spans a desired function
space~\cite{Woolrich2004}. More generally, one can also define a parametric
model of the HRF and estimate the parameters that best fit this
function~\cite{Lindquist2007}. However, in this case the estimated HRF may no longer be a linear function of the input parameters. 

Sensitivity to noise and overfitting can also be reduced through
regularization. For example, temporal regularization has been used in the
smooth FIR ~\cite{Goutte2000,Ciuciu2003,Casanova2008} to favor solutions with
small second order time derivative. These approaches require the setting of
one or several hyperparameters, at the voxel or potentially at the parcel
level (if several voxels in a pre-defined parcel are assumed to share some aspects of the HRF timecourse). Even if efficient techniques such as generalized   
\mbox{cross-validation}~\cite{golub1979generalized} can be used to choose the
regularization parameters, these methods are inherently more costly than 
\mbox{basis-constrained} methods. \mbox{Basis-constrained} methods also require
setting the number of basis elements; however, this parameter is not
continuous (as in the case of regularized methods), and in practice only few
values are explored: for example the 3-element basis set formed by a reference HRF
plus derivatives and the FIR model.  This paper focuses on basis-constrained
regularization of the HRF to avoid dealing with hyperparameter selection with
the goal of remaining computationally attractive.  A different approach to
increase robustness of the estimates consists in linking the estimated HRFs
across a predefined brain parcel, taking advantage of the spatially dependent nature of
fMRI~\cite{Wang2013}. However, \mbox{hemodynamically-informed}
parcellations~\cite{Chaari2012,Badillo2013a} rely on the computation of 
a large number of estimations at the voxel or \mbox{sub-parcel} level.
In this setting, the development of voxel-wise estimation procedures is complementary to the
development of parcellation methods in that more robust estimation
methods at the voxel level would naturally translate into more 
robust parcellation methods. In this paper we focus on voxel-wise
estimation methods.

We propose a method for the simultaneous estimation of HRF and
activation coefficients based on low-rank modeling. Within this
model, and as in~\cite{Makni2008,Kay2008,vincent2010spatially,Degras2014}, the HRF is constrained
to be equal across the different conditions, yet
permitting it to be different across voxels. 
Unlike previous works, we formulate this 
model as a constrained least squares problem, where the vector of coefficients
is constrained to lie within the space of rank one matrices. We
formulate the model within the framework of smooth optimization 
and use quasi-Newton methods to find the vector of estimates. 
This model was briefly presented
in the conference paper~\cite{Pedregosa2013}. Here we provide more
experimental validation and a more detailed presentation of the method. We also added results using a GLM with
separate designs~\cite{Mumford2012}.
Ten alternative approaches are now compared on two publicly available datasets. The solver has also been significantly improved to scale to full
brain data.

The contributions of this paper are two-fold.
First, we quantify the importance of HRF
estimation in encoding and decoding models. While the benefit of data-driven
estimates of the HRF have already been
reported in the case of decoding~\cite{Turner2012} and encoding
approaches~\cite{vu2011encoding}, we here provide a comprehensive
comparison of models. Second, we evaluate
a method called {\it GLM with Rank-1 constraint (R1-GLM)} that improves
encoding and decoding scores over \mbox{state-of-the-art} methods
while remaining computationally tractable on a full brain volume. We propose an
efficient algorithm for this method and discuss practical issues such as
initialization. Finally, we provide access to an open source software implementation of
the methods discussed in this paper.

\medskip \textbf{Notation:} $\|\cdot\|$ and $\|\cdot\|_{\infty}$ denote the
Euclidean and infinity norm for vectors. We use lowercase boldface letter to denote vectors and uppercase boldface letter to denote matrices. $\B{I}$ denotes the
identity matrix, $\mathbf{1}_n$ denotes the vector of ones of size $n$, $\otimes$ denotes
the Kronecker product and \(\vecop(\B{A})\) denotes the concatenation of the
columns of a matrix \(\B{A}\) into a single column vector. $\B{A}^{\dagger}$ denotes
the Moore-Penrose pseudoinverse. Given the vectors $\{\B{a}_1, \ldots, \B{a}_k \}$ with $\B{a}_i \in \RR^n$ for each $1 \leq i \leq k$, we will use the notation $[\B{a}_1, \ldots, \B{a}_k] \in \RR^{n \times k}$ to
represents the columnwise concatenation of the $k$ vectors into a matrix of size ${n \times k}$.  We will use Matlab-style colon notation to denote slices of an array, that is $\B{x}(1:k)$ will denote the first $k$ elements of $\B{x}$.

\section{Methods}

In this section we describe different methods for extracting the HRF and
activation coefficients from BOLD signals. We will refer to each different
stimulus as {\it condition} and we will call {\it trial} a unique presentation
of a given stimulus. We will denote by $k$ the total number of stimuli, $\B{y} \in \RR^n$
the BOLD signal at a single voxel and $n$ the total number of images acquired.

\subsection{The General Linear Model}

The original GLM model~\cite{Friston1995} makes the assumption that
the hemodynamic response is a linear transformation of the underlying neuronal
signal. We define the {$n \times k$}-matrix $\B{X}_{\text{GLM}}$ as the columnwise
stacking of different regressors, each one defined 
as the convolution of a reference HRF~\cite{Boynton1996,Glover1999} with the
stimulus onsets for the given condition.
In this work we used as reference HRF the one provided
by the software SPM 8~\cite{friston2011statistical}.
Assuming additive white noise, $n \geq k$ and $\B{X}_{\text{GLM}}$
 to be full rank, the vector of
estimates is given by $\hat{\bfbeta}_{\text{GLM}} =
{\B{X}_{\text{GLM}} ^ {\dagger}} \B{y}$, where $\hat{\bfbeta}_{\text{GLM}}$ is a vector of size
$k$ representing the amplitude of each one of the conditions in a given voxel.

A popular modification of this setting consists in extending the GLM design
matrix with the temporal and width derivatives of the reference HRF. This
basis, formed by the reference HRF and its derivatives with respect to time
and width parameters, will be used throughout this work. We will refer to it as
the \emph{3HRF basis}. In this case, each one of the basis elements is convolved
with the stimulus onsets of each condition, obtaining a design matrix of size $n
\times 3 k$. This way, for each condition, we estimate the form of the HRF as
a sum of basis functions that correspond to the first order Taylor expansion
of the parametrization of the response function. Another basis set that will be used is the Finite
Impulse Response (FIR) set. This basis set spans the complete ambient vector space (of dimension corresponding to the length of the impulse response) and it is thus a flexible model for capturing the HRF shape.
It consists of the canonical unit vectors (also known as stick function) for the given duration of the estimated
HRF. Other basis functions such as FMRIB's Linear Optimal Basis
Sets~\cite{Woolrich2004} are equally possible but were not considered in this
work.

More generally, one can extend this approach to any set of basis functions.
Given the matrix formed by the stacking of $d$ basis elements $\B{B} = [\B{b}_1, \B{b}_2, \ldots, \B{b}_d]$, the design matrix
$\B{X}_\B{B}$ is formed by successively stacking the regressors obtained by convolving
each of the basis elements with the stimulus onsets of each condition.
This results in a matrix of
size $n \times d k$ and under the aforementioned conditions
the vector of estimates is given
by
{$\hat{\bfbeta}_{\B{B}} = \B{X}_{\B{B}} ^ {\dagger} \B{y} $.} 
In this case, 
$\hat{\bfbeta}_{\B{B}}$ is no longer a vector of size $k$: it has length $k \times d$ instead
and can no longer be interpreted as the amplitude of the activation. One
possibility to recover the trial-by-trial reponse amplitude is to select the parameters from a single time point as done by 
some of the models considered in~\cite{Mumford2012}, however this procedure
assumes that the peak BOLD response is located at that time point. Another
possibility is to construct the estimated HRF and take as amplitude
coefficient the peak amplitude of this estimated HRF. This is the approach that
we have used in this paper.

% XXX : would it make sense to write vectors and matrices in bold ?
% So here \bfbeta_i with a FIR it is obvious that it is a vector.

\subsection{GLM with rank constraint}

In the basis-constrained GLM model, the HRF estimation is performed 
independently for each condition. This method works reliably whenever
the number of conditions is small, but in experimental designs with a large
number of conditions it performs poorly due to the limited conditioning of the problem and the increasing variance of the estimates.
% put ref?

At a given voxel, it is expected that for similar stimuli the estimated HRF are also 
similar~\cite{Henson2002}. Hence, a natural idea is to promote a common HRF
across the various stimuli (given that they are sufficiently similar), which should result in more robust estimates~\cite{Makni2008,vincent2010spatially}.
In this work we consider a model in which a common HRF is shared
across the different stimuli. Besides the estimation of the HRF,
a unique coefficient is obtained per column of our event
matrix. This amounts to the estimation of $k + d$ free parameters
instead of $k \times d$ as in the standard basis-constrained GLM setting.

The novelty of our method stems from the observation that the formulation of the GLM model with a
common HRF across conditions translates to a rank constraint on the vector of estimates. 
This assumption amounts to enforcing the vector of
estimates to be of the form $\bfbeta_{\B{B}} = [\mathbf{h} {\beta}_1, \mathbf{h} \beta_2, \cdots, \mathbf{h}
\beta_k]$ for some HRF $\mathbf{h} \in \RR^d$ and a vector of coefficients $\bfbeta \in \RR^k$. More compactly, this can be written as $\bfbeta_{\B{B}} = \vecop(\B{h}
\bfbeta^T)$. This can be
seen as a constraint on the vector of coefficients to be the vectorization of a rank-one
matrix, hence the name {\it Rank-1 GLM (R1-GLM)}.

In this model, the coefficients have no longer a closed form expressions,
but can be estimated by minimizing the mean squared error of a bilinear model. Given $\B{X}_{\B{B}}$ and $\B{y}$ as before, $\B{Z} \in \RR^{n \times q}$ a matrix of nuisance parameters such as drift regressors, we define $F_{\text{R1}}(\B{h}, \bfbeta, \bfomega, \B{X}_{\B{B}}, \B{y}, \B{Z}) = \frac{1}{2}\|\mathbf{y} - \mathbf{X}_{\B{B}} \vecop(\B{h} \bfbeta^T) - \B{Z} {\bfomega}\| ^2$ to be the objective function to be minimized. The optimization problem reads:
\begin{eqnarray}
\label{eq:r1}
\begin{aligned}
\hat{\B{h}}, ~\hat{\bfbeta},~ \hat{\bfomega} ~=~& \argmin_{\B{h}, \bfbeta, {\bfomega}} ~ F_{\text{R1}}(\B{h}, \bfbeta, \bfomega, \B{X}_{\B{B}}, \B{y}, \B{Z})\\
&\text{subject to } \|\B{B} \B{h}\|_{\infty} = 1 \text{ and } \langle \B{B} \B{h}, \B{h}_{\text{ref}}\rangle > 0 \enspace ,
\end{aligned}
\end{eqnarray}
The norm constraint is added to avoid the scale ambiguity between $\B{h}$ and $\bfbeta$
and the sign is chosen so that the estimated HRF correlates
positively with a given reference HRF $\B{h}_{\text{ref}}$.
Otherwise the signs of the HRF and $\bfbeta$ can be simultaneously flipped without changing
the value of the cost function. Within its feasible set, the optimization problem
is {\it smooth} and is convex with respect to $\B{h}$, $\bfbeta$ and $\bfomega$,
 however it is not {\it jointly convex} in variables $\B{h}$, $\bfbeta$ and $\bfomega$.

From a practical point of view this formulation has a number of advantages.
First, in contrast with the GLM without rank-1 constraint the estimated
coefficients are already factored into the estimated HRF and the activation
coefficients. That is, once the estimation of the model parameters
from Eq.~\eqref{eq:r1} is obtained, $\hat{\bfbeta}$ is a vector of size $k$ and $\hat{\B{h}}$ is a
vector of size $d$ that can be both used in subsequent analysis, while in models
without rank-1 constraint only the vector of coefficients (equivalent to 
$\text{vec}(\B{h} \bfbeta^T)$ in rank-1 constrained models) of size $k
\times d$ is estimated. In the latter case, the estimated HRF and the beta-maps
still have to be extracted from this vector by methods such as normalization by the peak of the HRF,
averaging or projecting to the set of Rank-1 matrices.

Second, it is readily adapted to prediction on unseen trials. While for
classical (non rank-1 models) the HRF estimation is performed per condition with no HRF associated with unseen conditions, in this setting, because the
estimated HRF is linked and equal across conditions it is natural to use this
estimate on unseen conditions. This setting occurs often in encoding models 
where prediction on unseen trials is part of the cross-validation procedure.

This model can also be extended to a parametric HRF model. That is,
given the hemodynamic response defined as a function $h: \RR^{d_1} \to \RR^d$ of some parameters
$\bfalpha$, we can formulate the analogous model of Eq.~\eqref{eq:r1} as an
optimization over the parameters $\bfalpha$ and $\bfbeta$ with the design matrix
$\B{X}_{\text{FIR}}$ given by the convolution of the event matrix with the FIR basis:
\begin{eqnarray}
\label{eq:r1_parametric}
\begin{aligned}
\hat{\bfalpha}, ~\hat{\bfbeta}, ~\hat{\bfomega} ~=~&\argmin_{\bfalpha, \bfbeta, \bfomega} 
F_{\text{R1}}(h(\bfalpha), \bfbeta, \bfomega, \B{X}_{\text{FIR}}, \B{y}, \B{Z}) \\
&\text{subject to } \| h(\bfalpha)\|_{\infty} = 1  \text{ and } \langle h(\bfalpha), \B{h}_{\text{ref}} \rangle > 0
\end{aligned}
\end{eqnarray}

In section \ref{sub:optim} we will discuss optimization strategies for both
models.

\subsection{Extension to separate designs}

An extension to the classical GLM that improves the estimation with correlated
designs was proposed in~\cite{Mumford2012}.
In this setting, each voxel is modeled as a linear combination of two
regressors in a design matrix $\B{X}_\text{GLM}$. The first one is the
regressor associated with a given condition and the second one is the sum of all
other regressors. This results in $k$ design matrices, one for each condition.
The estimate for a given condition is given by the first element in the two-dimensional
array ${\B{X}_{\text{S}i}}^{\dagger} \B{y}$, 
where $\B{X}_{\text{S}i}$ is the design matrix for
condition $i$. We will 
denote this model GLM with separate designs (GLMS). It has been reported to find
a better estimate in rapid event designs leading to a boost in accuracy for
decoding tasks~\cite{Mumford2012, Schoenmakers2013, Lei2013}.

This approach was further extended in \cite{Turner2012} to
include FIR basis instead of the predefined canonical function. Here we employ it 
in the more general setting of a  
predefined basis set. Given a set of basis
functions we construct the design matrix for condition $i$ as the columnwise
concatenation of two matrices $\B{X}^0_{\text{BS}i}$ and $\B{X}^1_{\text{BS}i}$.
$\B{X}^0_{\text{BS}i}$ is given by the columns associated
with the current condition in the GLM matrix and $\B{X}^1_{\text{BS}i}$ is
the sum of all other columns.
In this case, the vector of estimates is given by the first $d$ vectors of
$\B{X}_{\text{BS}i}^{\dagger} \B{y}$. See~\cite{Turner2012} for a more complete description of the matrices $\B{X}^0_{\text{BS}i}$ and $\B{X}^1_{\text{BS}i}$.

It is possible to use the same rank-1
constraint as before in the setting of separate designs, linking the HRF 
across conditions. We will refer to this model as \emph{Rank-1 GLM with separate designs (R1-GLMS)}. In this case the objective function has the form
$F_{\text{R1-S}}(\B{h}, \bfbeta, \bfomega, \B{r}, \B{X}_{\B{B}}, \B{y}, \B{Z}) = \frac{1}{2}\sum_i^k \|\B{y} - \beta_i \B{X}^0_{\text{BS}i} \B{h} - r_i \B{X}^1_{\text{BS}i} \B{h} - \B{Z} \bfomega\| ^2 $, where $\B{r} \in \RR^d$ is a vector representing the activation of all events except the event of interest and 
will not be used in subsequent analyses. We can 
compute the vector of estimates $\hat{{\bfbeta}}$ as the solution to the optimization
problem
\begin{eqnarray}
\label{eq:r1_separate}
\begin{aligned}
\hat{\bfbeta}, ~\hat{\bfomega},~\hat{\B{h}},~\hat{\bf{r}} ~= ~&\argmin_{\B{h}, \bfbeta, \bfomega, \B{r}} F_{\text{R1-S}}(\B{h}, \bfbeta, \bfomega, \B{r}, \B{X}_{\B{B}}, \B{y}, \B{Z}) \\
&\text{subject to } \|\B{B} \B{h}\|_{\infty} = 1 \text{ and } \langle \B{B} \B{h}, \B{h}_{\text{ref}} \rangle > 0
\end{aligned}
\end{eqnarray}

\subsection{Optimization}
\label{sub:optim}

For the estimation of rank-1 models on a full brain volume, a  model is estimate at each voxel separately. Since a typical brain volume contains more than 40,000 voxels, the efficiency of the estimation at a single voxel is of great importance. In this section we will detail an efficient procedure based on quasi-Newton methods for the estimation of R1-GLM and R1-GLMS models on a given voxel.

One approach to minimize \eqref{eq:r1} is to alternate the minimization
with respect to the variables $\bfbeta$, $\B{h}$ and $\bfomega$. By recalling the Kronecker product identities~\cite[Chapter 4.3]{horn1991topics}, and using the identity \(\vecop(\B{h}\bfbeta^T) = \bfbeta\otimes \B{h}\)
we can rewrite the objective function~\eqref{eq:r1} to be minimized as:
\begin{gather}
\label{eq:kron}
\frac{1}{2}\|\B{y} - \B{X}_{\B{B}} (\bfbeta \otimes \B{h}) - \B{Z} \bfomega\| ^2 = \\ \frac{1}{2}\|\B{y} - \B{X}_{\B{B}} (\B{I} \otimes \B{h}) \bfbeta - \B{Z} \bfomega\| ^2 = \\ \frac{1}{2}\|\B{y} - \B{X}_{\B{B}} (\bfbeta \otimes \B{I}) \B{h} - \B{Z} \bfomega\|^2 \enspace.
\end{gather}
Updating $\B{h}$, $\bfbeta$ or $\bfomega$ sequentially thus amounts to solving a (constrained) least squares
problem at each iteration. A similar procedure is detailed in~\citep{Degras2014}. However, this approach requires computing the
matrices $\B{X}_{\B{B}} (\bfbeta \otimes \B{I})$ and $\B{X}_{\B{B}} (\B{I} \otimes \B{h})$ at each iteration, which are typically dense,
resulting in a high computational cost per iteration. Note also that the optimization problem is not jointly convex in variables $\B{h}, \bfbeta, \bfomega$, therefore we cannot apply convergence guarantees from convex analysis.

We rather propose a more efficient approach by optimizing both variables jointly. We define a 
global variable $\B{z}$ as the concatenation of $(\B{h}, \bfbeta, \bfomega)$ into a single vector, $\B{z} = \vecop([\B{h}, \bfbeta, \bfomega])$,
 and cast the problem as an optimization with respect to this new variable.
Generic solvers for numerical
optimization~\cite{nocedal2006numerical} can then be used. The solvers that we will consider take as
input an objective function and its gradient. In this case, the partial derivatives with respect to variable $\B{z}$ can be written as 
$\partial F_{\text{R1}} / \partial \B{z} = \vecop([\partial F_{\text{R1}} / \partial \B{h}, \partial F_{\text{R1}} / \partial{\bfbeta}, \partial F_{\text{R1}} / \partial {\bfomega}])$, whose expression can be
easily derived using the aforementioned Kronecker product identities:
\begin{equation*}
    \left\{
    \begin{aligned}
        \frac{\partial F_{\text{R1}}}{\partial \B{h}}=& - (\bfbeta^T \otimes \B{I}) \B{X}^T (\B{y} - \B{X} \vecop(\B{h} \bfbeta^T) - \B{Z} \bfomega) \\
        \frac{\partial F_{\text{R1}}}{\partial \bfbeta}=& - (\B{I} \otimes \B{h}^T) \B{X}^T (\B{y} - \B{X} \vecop(\B{h} \bfbeta^T) - \B{Z} \bfomega) \\
        \frac{\partial F_{\text{R1}}}{\partial \bfomega}=& - \B{Z}^T (\B{y} - \B{X} \vecop(\B{h} \bfbeta^T) - \B{Z} \bfomega)
    \end{aligned}
    \right.
\end{equation*}

If instead a parametric model of the HRF is used as in Eq.~\eqref{eq:r1_parametric}, the equivalent partial derivatives can be easily computed by the chain rule.

For the sake of efficiency, it is essential to avoid evaluating the Kronecker products naively,
 but rather reformulate them using the above mentioned Kronecker identities. For example, the matrix $\B{M} = \B{X} (\B{I} \otimes \B{h})$ should not be computed explicitly but should rather be stored as a linear operator such that when applied to a vector $\bfbeta \in \RR^k$ it computes $M(\bfbeta) = \B{X} (\bfbeta \otimes \B{h})$, avoiding thus the explicit computation of $\B{I} \otimes \B{h}$. 

 % Furthermore, $\B{X} (\B{a} \otimes \B{h})$ can be computed as $\B{a}^T \tilde{\B{M}} \B{h}$ for $\tilde{\B{M}}$ a reshape of $\B{M} \RR^{n \times p r}$ into an array of shape $p \times $

Similar equations can be derived for the rank-1 model with separate designs of Eq.~\eqref{eq:r1_separate} (\mbox{R1-GLMS}), in which
case the variable $\B{z}$ is defined as the concatenation of $(\B{h}, \bfbeta, \bfomega, \B{r})$, i.e. $\B{z} = \vecop([\B{h}, \bfbeta, \bfomega, \B{r}])$. The gradient of $F_{\text{R1-S}}$ with respect to $\B{z}$ can be computed as $\partial F_{\text{R1-S}} / \partial \B{z} = \vecop([\partial F_{\text{R1-S}} / \partial \B{h}, \partial F_{\text{R1-S}} / \partial{\bfbeta}, \partial F_{\text{R1-S}} / \partial {\bfomega}, F_{\text{R1-S}} / \partial {\B{r}}])$. The partial derivatives read:
\begin{equation*}
    \left\{
    \begin{array}{lcl}
        \frac{\partial F}{\partial \B{h}} &=& \sum_i^k - (\B{X}^0_{\text{BS}_i}\bfbeta_i - \B{X}^1_{\text{BS}_i} r_i)^T (\B{y} - \bfbeta_i \B{X}^0_{\text{BS}_i} h - w_i \B{X}^1_{\text{BS}_i} \B{h}) \\
        \frac{\partial F}{\partial \beta_i} &=& -(\B{X}^0_{\text{BS}_i} \B{h})^T (\B{y} - \bfbeta_i \B{X}^0_{\text{BS}_i} \B{h} - w_i \B{X}^1_{\text{BS}_i} \B{h}) \\
        \frac{\partial F}{\partial \omega_i} &=& -\B{Z}^T (\B{y} - \bfbeta_i \B{X}^0_{\text{BS}_i} \B{h} - w_i \B{X}^1_{\text{BS}_i} \B{h}) \\
        \frac{\partial F}{\partial r_i} &=& -(\B{X}^1_{\text{BS}_i} \B{h})^T (\B{y} - \bfbeta_i \B{X}^0_{\text{BS}_i} \B{h} - w_i \B{X}^1_{\text{BS}_i} \B{h}) \\
    \end{array}
    \right.
\end{equation*}

% \todo[color=blue!60, inline]{Used an eqnarray*, because this was awfully "align*"ed. It still isn't perfect though. Maybe we need to bind the left hand sides to the left? The same theoretically goes for the equations just above}

A good initialization plays a crucial role in the convergence of any iterative
algorithm. Furthermore, for non-convex problems a good initialization prevents the algorithm from converging to undesired local minima. We have used as initialization for the R1-GLM and R1-GLMS models the solution given by the GLM with
separate designs (GLMS). Since the GLM with separate designs scales linearly in the number of voxels, this significantly reduces computation time whenever
an important number of voxels is considered.

Whenever the design matrix $\B{X}_{\B{B}}$ has more rows than columns (as is
the case in both datasets we consider with $\B{B}$ the 3HRF basis), it is possible to
find an orthogonal transformation that significantly speeds up the computation
of the Rank-1 model. Let $\B{Q}, \B{R}$ be the ``thin'' QR decomposition of
$\B{X}_{\B{B}} \in \RR^{n \times d k}$, that is, $\B{Q} \B{R} = \B{X}_{\B{B}}$ with $\B{Q}
\in \RR^{n \times d k}$ an orthogonal matrix and $\B{R} \in \RR^{d k \times d k}$ 
a triangular matrix. Because of the invariance of the Euclidean norm to orthogonal
transformations, the change of variable $\B{X}_{\B{B}} \leftarrow \B{Q}^T
\B{X}_{\B{B}}$, $\B{y} \leftarrow \B{Q}^T \B{y}$ yields a Rank-1 model in Eq.~\eqref{eq:r1}
with equivalent solutions. This reduces the size of the design matrix to a square triangular matrix of size $d k \times d k$ (instead of $n \times d k$) and reduces the explained variable $\B{y}$ to a vector of size $k d$ (instead of $n$). After this change of variable, the convergence of the Rank-1 model is significantly faster due to the faster computation of the objective function and its partial derivatives.  We have observed that the total running time of the algorithm can be
reduced by 30\% using this transformation.

Some numerical solvers such as L-BFGS-B~\cite{liu1989limited}
require the constraints to be given as box constraints. While our original
problem includes an equality constraint we can easily
adapt it to use convex box constraints instead.
We replace the equality constraint $\|\B{B h}\|_{\infty} = 1$ by
the convex inequality constraint $\|\B{B h}\|_{\infty} \leq 1$, which is equivalent
to the box constraint $-1 \leq (\B{B h})_i \leq 1$ supported by the above solver. 
However, this change of constraint
allows solutions in which $\B{h}$ can be arbitrarily close to zero. To avoid such
degenerate cases we add the smooth term $-\|\B{B}(:, 1) h_1 \|^2 _2$ to the cost function. Since
there is a free scale parameter between $\B{h}$
and $\bfbeta$, this does not bias the problem, but forces $\B{B h}$ to lie as far as possible from the origin (thus saturating the box constraints). Once a descent
direction has been found by the \mbox{L-BFGS-B} method we perform a line search
procedure to determine the step length. The line-search
procedure was implemented to satisfy the strong Wolfe conditions~\cite{nocedal2006numerical}.
Finally, when the optimization algorithm has converged to a stationary point, 
we rescale the solution setting to ensure that the equality constraint. This still leaves a sign ambiguity between the estimated HRF and the associated beta-maps. To make these parameters identifiable, the sign of the estimated HRF will be chosen so that these correlate positively with the reference HRF.

We have compared several first-order (Conjugate Gradient), \mbox{quasi-Newton}
(L-BFGS) and Newton methods on this problems and found that in general \mbox{quasi-Newton} methods performed best in terms of computation time. In our
implementation, we adopt the L-BFGS-B as the default solver. 

In Algorithm~\ref{alg1} we describe an algorithm based on L-BFGS that can be used to optimize R1-GLM and R1-GLMS models (a reference implementation for the Python language is described in subsection Software). Variable $\B{r}$ is only used for the R1-GLMS method and its use is denoted within parenthesis, i.e. $(, \B{r})$, so that for the R1-GLM it can simply be ignored.
\begin{algorithm}
\caption{Optimization of R1-GLM and R1-GLMS models}
\label{alg1}
\begin{algorithmic}[1]
\REQUIRE Given initial points $\bfbeta_0 \in \RR^k, \B{h}_0 \in \RR^d, {\bfomega}_0 \in \RR^q ~(, \mathbf{r}_0 \in \RR^k)$, convergence tolerance $\epsilon > 0$, inverse Hessian approximation $\B{H}_0$. 
\ENSURE $\bfbeta_m, \B{h}_m$
\STATE {(Optional)}: Compute the QR decomposition of $\B{X}_{\B{B}}$, $\B{Q} \B{R} = \B{X}_{\B{B}}$, and replace $\B{X}_{\B{B}} \leftarrow \B{Q}^T \B{X}_{\B{B}}, \B{y} \leftarrow \B{Q}^T \B{y}$
\STATE Initialization. Set $m \leftarrow 0$, $\B{z} \leftarrow \vecop([\B{h}_0, \bfbeta_0, \bfomega_0 (, \B{r}_0)])$
\WHILE{$\|\nabla f\| > \epsilon$}
\STATE Compute search direction.  Set $\B{p}_m \leftarrow - \B{H}_m \nabla f(\B{h}_m, \bfbeta_m, \bfomega_m (, \B{r}_m))$
by means of the L-BFGS algorithm.
\STATE {Set $\B{z}_{m+1} = \B{z}_m + \gamma_m \B{p}_m$, where $\gamma_m$ is computed from a line search procedure subject to the box constraints $\|\B{h}_m\|_{\infty} \leq 1$}.
\STATE $m \leftarrow m+1$
\ENDWHILE 
\STATE Extract R1-GLM(S) parameters from $\B{z}_m$. Set $\B{h}_m \leftarrow \B{z}_m(1:d), \bfbeta_m \leftarrow \B{z}_m(d+1:m+d)$
\STATE Normalize and set sign so that the estimated HRF is positively correlated with a reference HRF: $q_m \leftarrow \|\B{h}_m\|_{\infty} \text{sign}(\B{h}_m^T \B{h}_{\text{ref}}),~ \B{h}_m \leftarrow \B{h}_m / q_m, ~\bfbeta_m \leftarrow \bfbeta_m q_m$
\end{algorithmic} 
\end{algorithm}

The full estimation of the R1-GLM model with 3HRF basis for one subject of the
dataset described in section {\it Dataset 2: decoding of potential gain
levels} ($16 \times 3$ conditions, $720$ time points, $41,622$ voxels) took 14
minutes in a 8-cores Intel Xeon 2.67GHz machine. The total running time for
the 17 subjects was less than four hours.

\subsection{Software}

We provide a software implementation of all the models discussed in this section
in the freely available (BSD licensed) pure-Python package \textsf{hrf\_estimation}
\footnote{\href{https://pypi.python.org/pypi/hrf\_estimation}{https://pypi.python.org/pypi/hrf\_estimation}}. 

\section{Data description}

With the aim of making the results in this paper easily reproducible, we have
chosen two freely available datasets to validate our approach and to compare
different HRF modeling techniques.

\subsection{Dataset 1: encoding of visual information}

The first dataset we will consider is 
described in~\cite{Kay2008,naselaris2009bayesian,kay2011data}. It contains 
BOLD fMRI responses in human subjects viewing natural images.
As in~\cite{Kay2008}, we performed
prediction of BOLD signal following the visual presentation of natural images
and compared it against the measured fMRI BOLD signal.
As the procedure consists of predicting the fMRI data
from stimuli descriptors, it is an {\it encoding} model.
This dataset is publicly available from \url{http://crcns.org}

Two subjects viewed 1750 training images, each presented twice,
and 120 validation images, each presented 10 times, while fixating a
central cross. Images were flashed 3 times per second (200 ms
on-off-on-off-on) for one second every 4 seconds, leading to a rapid
event-related design. The data were acquired in 5 scanner sessions on 5 different days,
each comprising 5 runs of 70 training images --each image  being presented twice
within the run-- and 2 runs of validation images showing 12 images,
10 times each. The images were recorded from the occipital cortex at a spatial resolution of 2mm$\times$2mm$\times$2.5mm
and a temporal resolution of 1 second. Every brain volume for each subject has been aligned to the first volume of the first run of the 
first session for that subject. Across-session alignment was performed manually. Additionally, 
data were temporally interpolated to account for slice-timing differences. See~\citep{Kay2008} for further preprocessing details. 

We performed local detrending using a Savitzky-Golay
filter~\cite{savitzky1964smoothing} with a polynomial of degree 4 and a window
length of 91 TR. The activation coefficients (beta-map) and HRF were extracted from
the training set by means of the different methods we would like to compare. The
training set consisted of 80\% of the original session (4 out of 5 runs). This
resulted in estimated coefficients (beta-map) for each of the $70 \times 4$
images in the training set.

We proceed to train the encoding model. The stimuli are handled as local image contrasts,  that are represented by spatially
smoothed Gabor pyramid transform modulus with 2 orientations and 4 scales. 
Ridge regression (regularization parameter chosen by Generalized Cross-Validation~\cite{golub1979generalized}) was then used to learn a
predictor of voxel activity on the training set. By using this encoding model
and the estimated HRF it is possible to predict the BOLD signal for the 70
images in the test set (20 \% of the original session). We emphasize that learning the HRF on the training set instead of on the full
dataset is necessary to avoid overfitting while assessing the quality of the estimated HRF by any
HRF-learning method: otherwise, the estimation of the HRF may incorporate specificities of the test set leading to artificially higher scores.

In a first step, we perform the image identification task from~\citep{Kay2008}. From the training set we estimate the activation coefficients that will be used to compute the activation maps. We use an encoding model using Gabor filters that predicts the activation coefficient from the training stimuli. From the stimuli in the validation set we predict the activation coefficients that we then use to identify the correct image. The predicted image is the one yielding the highest correlation with the measured activity. This procedure mimics the one presented in~\citep[Supplementary material]{Kay2008}.

In a second step, we report score as the Pearson correlation between the measurements and the  predicted BOLD signal on left out data. The prediction of BOLD signal on the test set is performed from conditions that
were not present in the train set. In order to do this, an HRF for these conditions is necessary. As highlighted in the methods section, the
construction of an HRF for these conditions is ambiguous for non Rank-1
methods that perform HRF estimation on the different stimuli. In these cases
we chose to use the mean HRF across conditions as the HRF for unseen
conditions. Finally, linear predictions on the left out fold were compared to
the measured BOLD signals.

\subsection{Dataset 2: decoding of potential gain levels}

The second dataset described in~\cite{Tom2007} is a gambling task where each
of the 17 subjects was asked to accept or reject gambles that offered a 50/50
chance of gaining or losing money. The magnitude of the potential gain and
loss was independently varied across 16 levels between trials. Each gamble has
an amount of potential gains and potential losses that can be used as class label. In
this experiment, we only considered gain levels. This leads to the challenge of
predicting or \emph{decoding} the gain level from brain images. The dataset
is publicly available from \url{http://openfmri.org} under the name 
\emph {mixed-gambles task} dataset.

The data preprocessing included slice timing, motion correction, coregistration to the
anatomical images, tissue segmentation, normalization to MNI space and was
performed using the SPM 8 software through the
Pypreprocess\footnote{\url{https://github.com/neurospin/pypreprocess}}
interface.

For all subjects three runs were recorded, each consisting of 240 images with
a repetition time (TR) of 2 seconds and a stimulus presentation at every 4
seconds. In order to perform HRF estimation on more data than what is
available on a single run, we performed the estimation on the three runs
simultaneously. This assumes HRF consistency across runs, which was obtained by
concatenating the data from the three runs and creating a block-diagonal design matrix correspondingly (each block is the design of one run).

After training a regression model on 90\% of the data, we predict the gain level on the
remaining 10\%. As a performance measure we use Kendall tau rank correlation
coefficient~\cite{kendall1938new} between the true gain levels and the
predicted levels, which is a measure for the orderings of the data.  We argue
that this evaluation metric is better suited than a regression loss for this
task because of the discrete and ordered nature of the labels. Also, this loss
is less sensible to shrinkage of the prediction that might occur when penalizing a
regression model~\citep{bekhti:hal-01032909}. The Kendall tau coefficient always lies within the interval
$[-1, 1]$, with $1$ being perfect agreement between the two rankings and $-1$
perfect disagreement. Chance level lies at zero. This metric was previously
proposed for fMRI decoding with ordered labels in~\cite{Doyle2013}.

\section{Results}

In order to compare the different methods discussed previously, we ran
the same encoding and decoding studies while varying the
estimation method for the activation coefficients (beta-maps). The
methods we considered are standard GLM (denoted GLM), GLM with separate
designs (GLMS), Rank-1 GLM (R1-GLM) and Rank-1 GLM with separate designs
(R1-GLMS). For all these models we consider different basis sets for
estimating the HRF: a set of three elements formed by the reference HRF and
its time and dispersion derivative, a FIR basis set (of size 20 in the
first dataset and of size 10 in the second dataset) formed by the canonical vectors
and the single basis set formed by the reference HRF (denoted ``fixed HRF''), which
in this case is the HRF used by the SPM 8 software.

It should be reminded that the focus of this study is not the study of the HRF
in itself (such as variability across subjects, tasks or regions) but instead
its possible impact on the accuracy of encoding and decoding paradigms. For
this reason we report encoding and decoding scores but we do not investigate
any of the possible HRF variability factors.

\subsection{Dataset 1: encoding of visual information}

In the original study, 500 voxels were used to perform image identification. These voxels were selected as the voxels with the highest correlation with the true BOLD signal on left-out data using a (classical) GLM with the reference HRF. These voxels are therefore not the ones naturally benefiting the most from HRF estimation.

\begin{figure} \centering
\includegraphics[width=1.\linewidth]{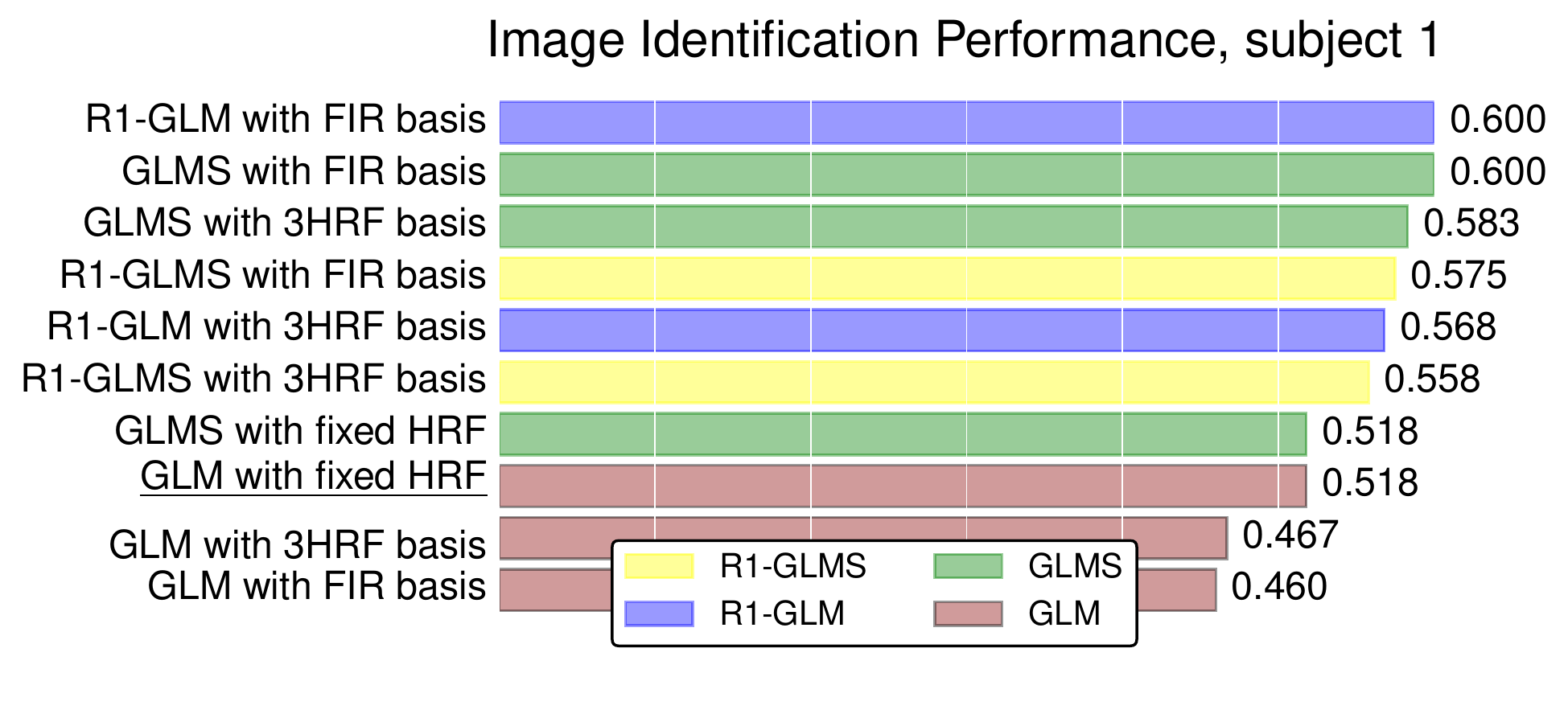}
\includegraphics[width=1.\linewidth]{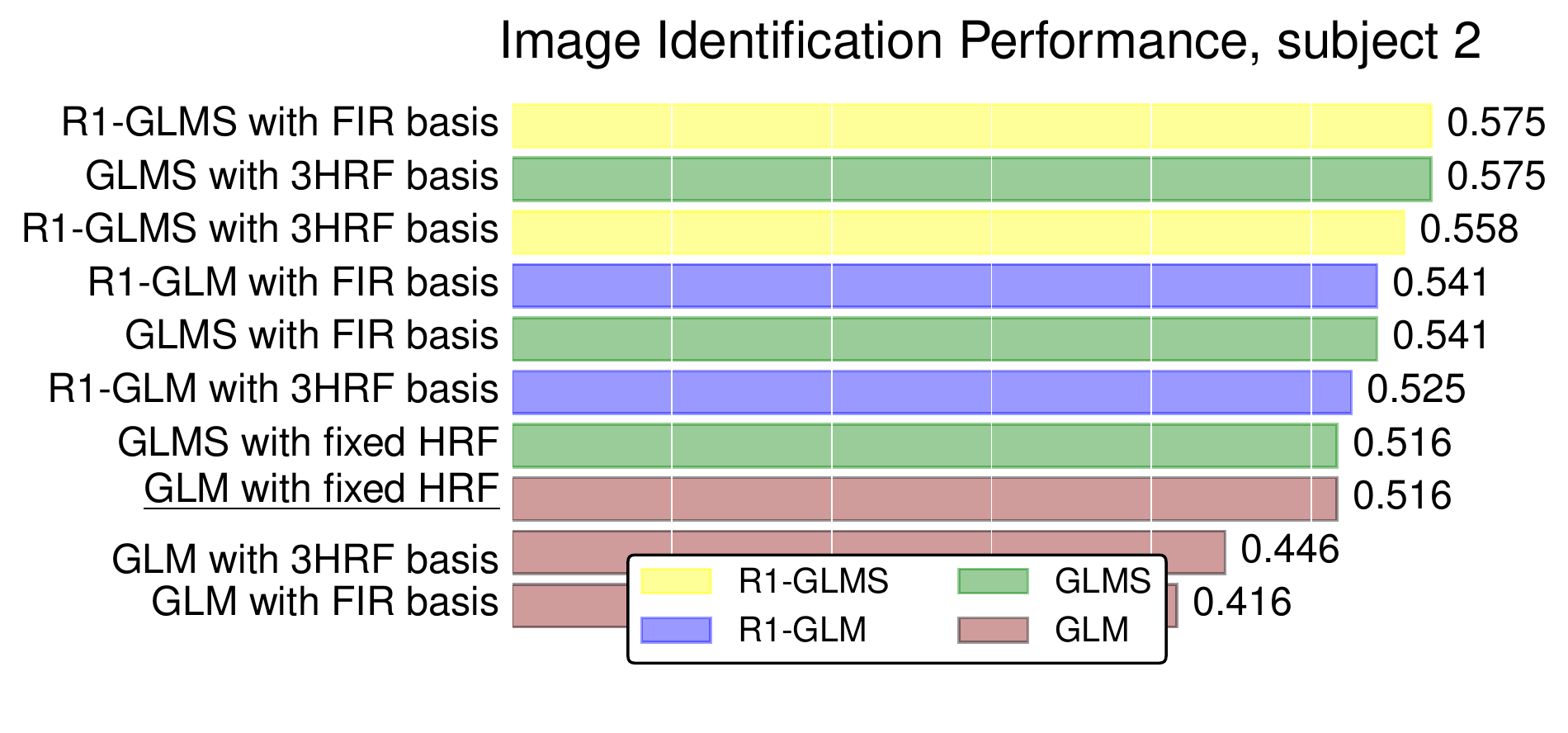}
\caption{\label{fig:identification_scores} Image identification score (higher is better) on two different subjects from the first dataset. The metric counts the number of correctly identified images over the total number of images (chance level is 1/120 $\approx 0.008$). This metric is less sensitive to the shape of the HRF than the voxel-wise encoding score. The benefits range from 0.9\% points to 8.2\% points across R1-constrained methods and subjects. The highest score is achieved by a R1-GLM method with a FIR basis set for subject 1 and by a R1-GLMS with FIR basis for subject 2.
}
\end{figure}

We first present the scores obtained in the image identification task for different variants of the GLM. This can be seen in Figure~\ref{fig:identification_scores}. The displayed score is the count of correctly identified images over the total number of images (chance level is therefore at 1/120). The identification algorithm here only uses the beta-maps obtained from the train and validation set. This makes the estimation of the HRF an intermediate result in this model. However, we expect that a correct estimation of the HRF directly translates into a better estimation of the activation coefficients in the sense of being able to acheive higher predictive accuracy. Our results are consistent with this hypothesis and in this task the rank-one (R1) and glm-separate (GLMS) models outperform the classical GLM model. The benefits range from 0.9\% for R1-GLM in subject 2 to 8.2\% for the same method and subject 1. It is worth noticing that methods with FIR basis obtain a higher score than methods using the 3HRF basis.

In order to test whether this increase is statistically significant we performed the following statistical test. The success of recovering the correct image can be modeled as a binomial distribution, with $p_A$ being be the probability of recovering the correct image with method A and $p_{{B}}$ be the probability of recovering the correct image with method B. We define the null hypothesis $H_0$ as the statement that both probabilities are equal, $H_0: p_A = p_{{B}}$, and the alternate hypothesis that both probabilities and not equal, $H_1: p_1 \neq p_2$ (this test is sometimes known as the binomial proportion test~\citep{rohmel1999unconditional}). The score test statistic for the one-tailed test is $T = {(p_A - p_{{B}})} / {\sqrt{p (1 - p)\frac{2}{n}}}$, where $p = (p_A + p_{{B}}) / 2$ and $n$ is the number of repetitions, in this case $n=120$. This statistic is normally distributed for large $n$. The p-value associated with this statistical test when comparing every model (by order of performance) with the model ``GLM with with fixed HRF'' is $(0.10, 0.10, 0.15, 0.19, 0.21, 0.26, 0.5, 0.5, 0.82, 0.81)$ for the first subject and $(0.18, 0.18, 0.25, 0.34, 0.34, 0.44, 0.5, 0.5, 0.86, 0.93)$ for the second.

\begin{figure} \centering
\includegraphics[width=1.\linewidth]{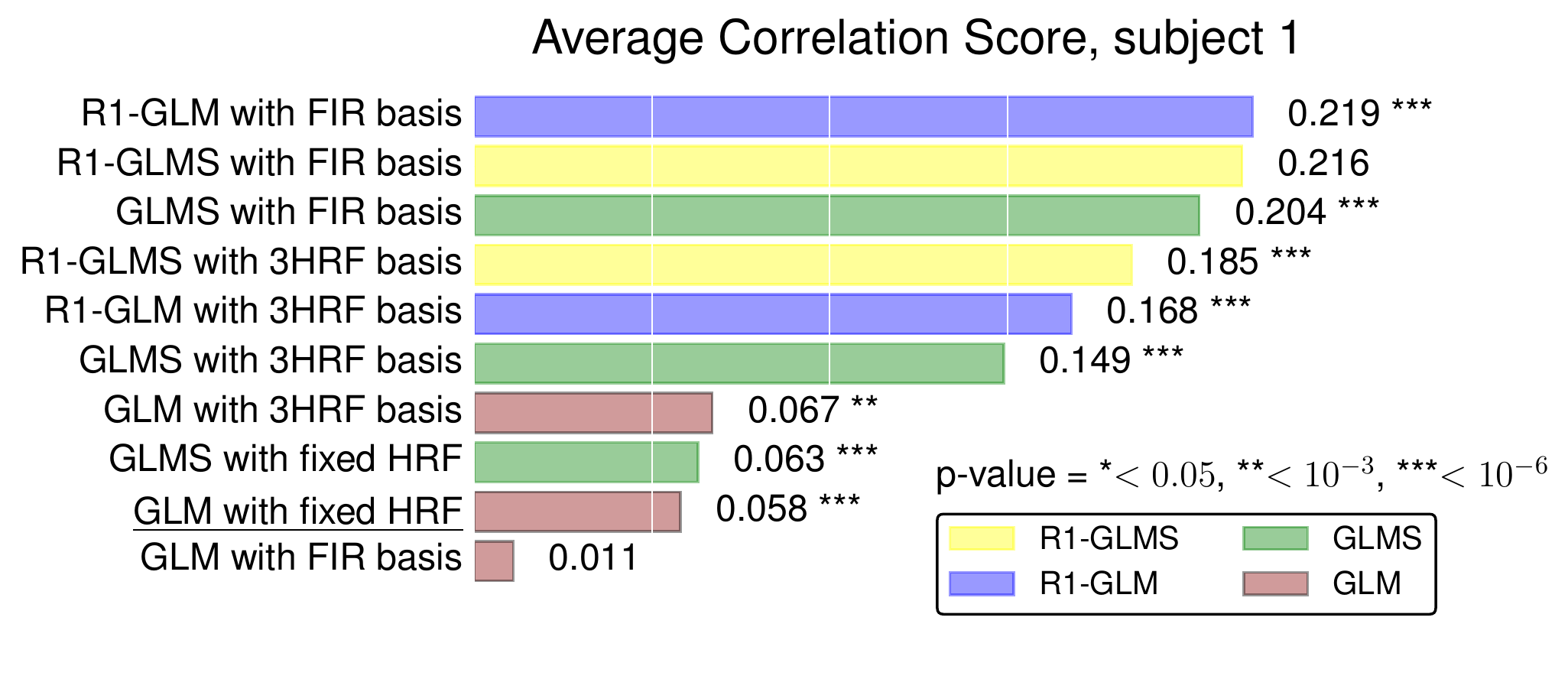}
\includegraphics[width=1.\linewidth]{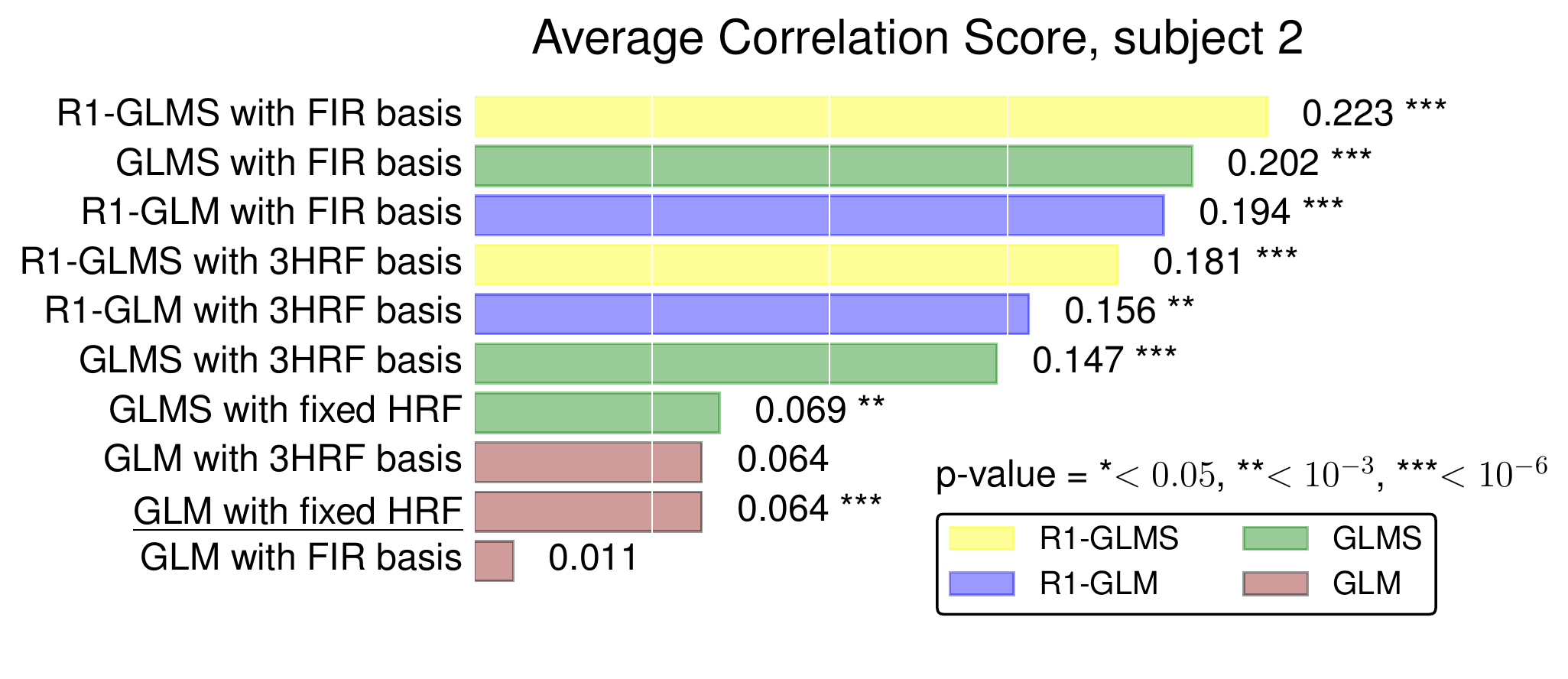}
\caption{\label{fig:encoding_scores} Average correlation score (higher is better) on two different subjects from the first dataset. The average correlation score is the Pearson correlation between the predicted BOLD and the true BOLD signal on left-out session, averaged across voxels and sessions. Methods that perform constrained HRF estimation significantly outperform 
methods that use a fixed reference HRF. As for the image identification performance,
the best performing method for subject 1 is the R1-GLM, while for subject 2 it is the R1-GLMS model, both with FIR basis. 
In underlined typography is the GLM with a fixed HRF which is the method
used by default in most software distributions.
A Wilcoxon signed-rank test is performed
between each method and the next one in the ordered result list by considering the leave-one-session out cross-validation scores for each method.
We report p-values to assess whether the score differences
are statistically significant.
}
\end{figure}

We will now use a different metric for evaluating the performance of the encoding model. This metric is the Pearson correlation between the BOLD predicted by the encoding model and the true BOLD signal, averaged across voxels. We will compute the this metric on a left-out session, which results in five scores for each method, corresponding to each of the cross-validation folds. Given two methods, a Wilcoxon signed-rank test can be used on these cross-validation scores to assess whether the score obtained by the two methods are significantly different. This way, irrespective of the  variance across voxels, which is inherent to the study, we can reliably  assess the relative ranking of the different models. In Figure~\ref{fig:encoding_scores} we show the scores for each method (averaged across sessions) and the p-value corresponding the Wilcoxon test between a given method and the previous one by order of performance.

We observed in Figure~\ref{fig:encoding_scores} that methods that learn the HRF
together with some sort of regularization (be it Rank-1 constraint or induced
by separate designs) perform noticeably better than methods that perform
unconstrained HRF estimation, highlighting the importance of a robust
estimation of the HRF as opposed to a free estimation as performed by the
standard GLM model with FIR basis. This suggests that R1 and GLMS methods permit
 including FIR basis sets while minimizing the risk of overfitting inherent to the classical GLM model.

We also observed that models using the GLM
with separate designs from~\cite{Mumford2012} perform significantly better on
this dataset than the standard design, which is consistent with the purpose of
these models. It improves estimation in highly correlated designs. The best
performing model for both subjects in this task is the R1-GLMS with FIR basis, followed by
the R1-GLM with FIR basis model for subject 1 and GLMS with FIR basis for
subject 2. The difference between both models (Wilcoxon signed-rank test) was
significant with a p-value $< 10^{-6}$. Since the results for both
subjects are similar, we will only use subject 1 for the rest of the figures.

\begin{figure}
\centering
\includegraphics[width=1.\linewidth]{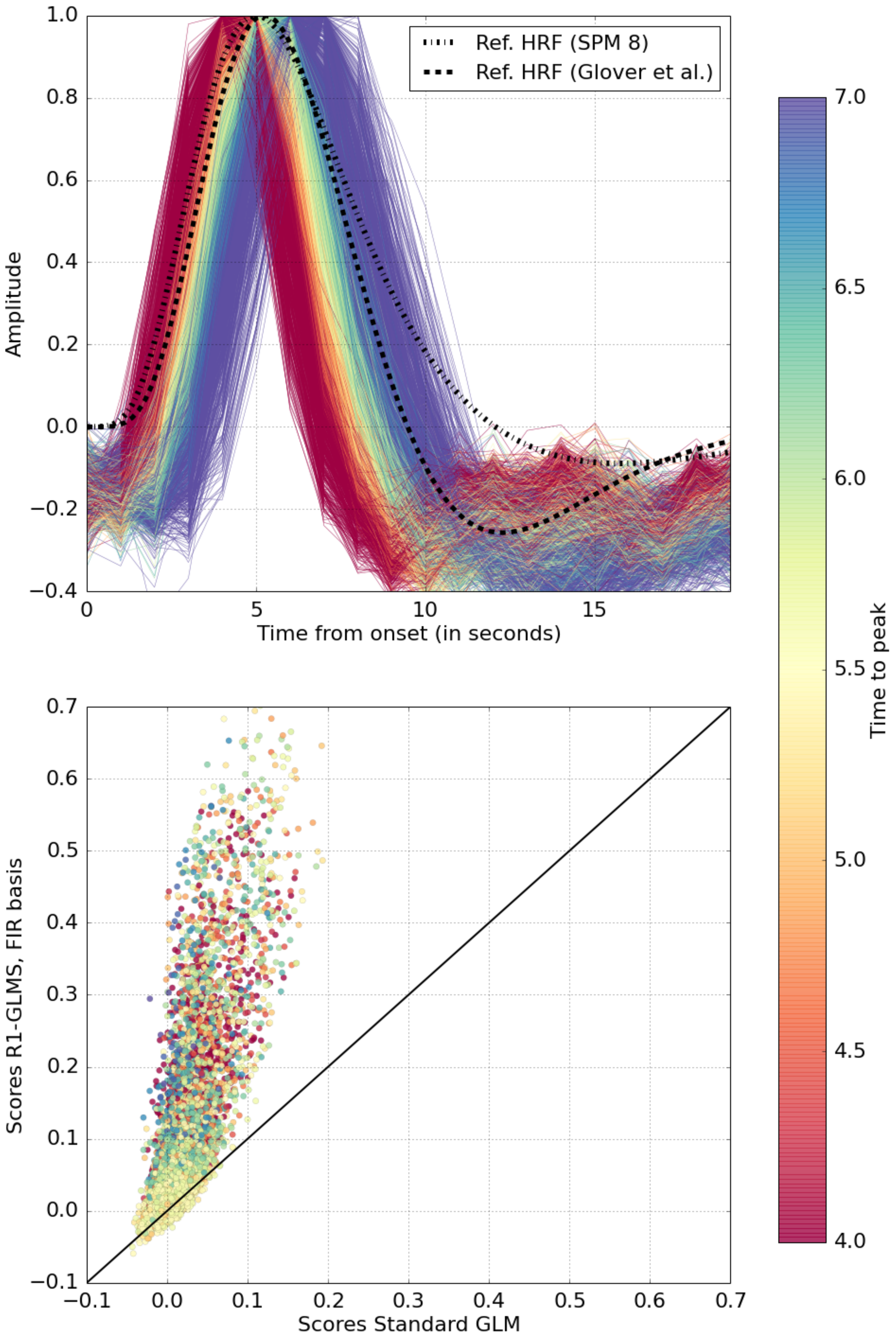}
\caption{\label{fig:r1_vs_can}
Top: HRF estimated by the R1-GLMS method on voxels for which the encoding score was
above the mean encoding score (first dataset), color coded according to the time to peak 
of the estimated HRFs. 
The difference in the estimated HRFs suggests a substantial variability at the voxel level 
within a single subject and a single task. Bottom: voxel-wise encoding score 
for the best
performing method (R1-GLMS with FIR basis) versus a standard GLM (GLM with fixed HRF)
across voxels. The metric is Pearson correlation. Points 
above the black diagonal correspond to voxels that exhibit a higher score with the R1-GLMS
method than with a standard GLM.
}
\end{figure}

To further inspect the results, we investigated the estimation and
encoding scores at the voxel level. This provides some valuable information.
For example, parameters such as time-to-peak, width and undershoot of the
estimated HRF can be used to characterize the mis-modeling of a reference HRF
for the current study. Also, a voxel-wise comparison of the different methods
can be used to identify which voxels exhibit a greater improvement for a given
method. In the upper part of Figure~\ref{fig:r1_vs_can} we show the HRF
estimated on the first subject by our best performing method (the Rank-1 with separate designs and
FIR basis). For comparison we also present two commonly used
reference HRFs: one used in the software SPM and one defined in~\cite[auditory
study]{Glover1999} and used by software such as
NiPy\footnote{\href{http://nipy.org}{http://nipy.org}}  and
fmristat\footnote{\href{http://www.math.mcgill.ca/keith/fmristat/}{http://www.math.mcgill.ca/keith/fmristat/} }. Because the HRF
estimation will fail on voxels for which there is not enough signal, we only
show the estimated HRF for voxels for which the encoding score is above the
mean encoding score. In this plot the time-to-peak of the estimated HRF is
color coded. One can observe a substantial variability in the time to peak,
confirming the existence of a non-negligeable variability
of the estimated HRFs, even within a single subject and a single task. In
particular, we found that only 50\% of the estimated HRFs on the full brain volume 
peaked between 4.5 and 5.5 seconds.

In the lower part of Figure~\ref{fig:r1_vs_can} we can see a scatter plot in which the
coordinates of each point are the encoding scores with two
different methods. The first coordinate (X-axis) is given by the score using a
canonical GLM whilst the second coordinate (Y-axis) corresponds to the Rank-1 separate
with FIR basis. Points above the black diagonal
exhibit a higher score with our method than with a canonical GLM. As
previously, the color represents the time to peak of the estimated HRF.
From this plot we can see that voxels that have a low correlation
score using a canonical GLM do not gain significant
improvement by using a Rank-1 Separate FIR model instead. However, voxels that
already exhibit a sufficiently high correlation score using a canonical
GLM ($> 0.05$) see a significant increase in performance when estimated using
our method.

% XXX : don't you mean the contrary?

These results suggest as a strategy to limit the computational cost of learning the HRF
on an encoding study to perform first a standard GLM (or GLMS) on the full
volume and then perform HRF estimation only on the best performing voxels.

\begin{figure} \centering
\includegraphics[width=1.\linewidth]{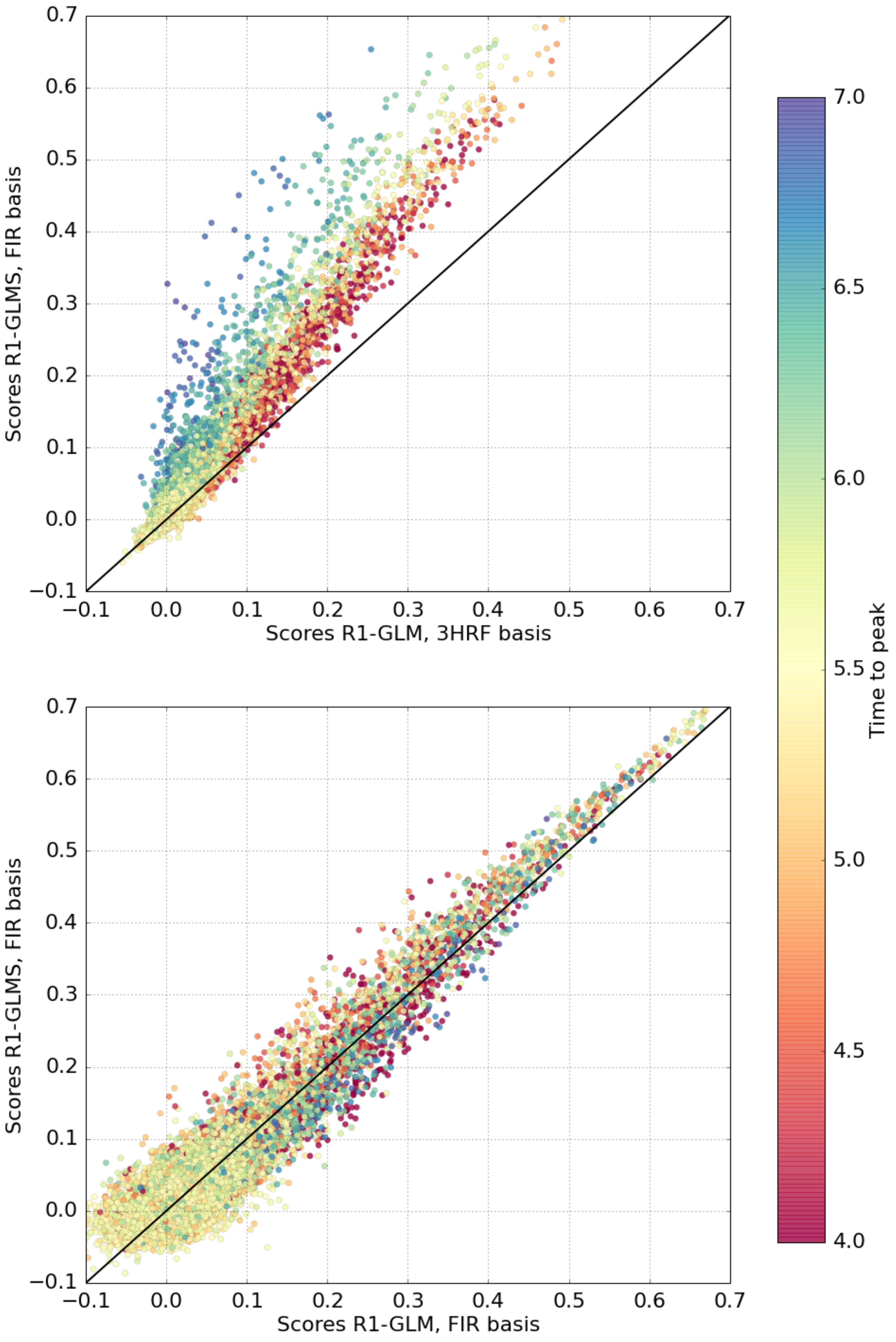}
\caption{\label{fig:scatter2} Voxel-wise encoding score for different models
that perform HRF estimation (first dataset). As in figure~\ref{fig:r1_vs_can},
color codes for the time to peak of the estimated HRF at the given voxel.
Top: two Rank-1 separate design models with different basis functions: FIR with 
20 elements in the Y-axis and the reference HRF with its time and dispersion derivatives
(3HRF) in the X-axis. The color trend in this plot suggests that the score improvement 
of the FIR basis with respect to the 3HRF
becomes more pronounced as the time-to-peak of the estimated HRF 
deviates from the reference HRF (peak at 5s). This can be explained by taking into account that the 3HRF
basis is a local model of the HRF around the peak time of the canonical HRF.
Bottom: voxel-wise encoding score for two Rank-1 models with FIR basis and 
different design matrices: separate design on the Y-axis and classical
design on the X-axis. Although both models give similar results, a Wilcoxon
signed-rank test on the leave-one-session-out cross-validation score (averaged across voxels) confirmed the superiority of the separate designs model
in this dataset with p-value $<10^{-3}$.}
\end{figure}

The methods that we have considered for HRF estimation can be subdivided according
to the design matrices they use (standard or separate) and the basis they use
to generate the estimated HRF (3HRF and FIR). We now focus on the performance
gains of each of these individual components.
In the upper part of
Figure~\ref{fig:scatter2} we consider the top-performing model, the Rank-1
GLMS, and compare the performance of two different basis sets: FIR with 
20 elements in the Y-axis and the reference HRF plus its time and dispersion derivatives
(3HRF) in the X-axis. The abundance of points above the diagonal 
demonstrates the superiority of the FIR basis on this dataset.
The color trend in this plot suggests that the score improvement of the FIR basis
with respect to the 3HRF basis
becomes more pronounced as the time-to-peak of the estimated HRF 
deviates from the reference HRF (peak at 5s), which can be explained by observing that 
the 3HRF basis corresponds to a local model around the time-to-peak. 
In the bottom part of this figure 
we compare the different design matrices (standard or separate). Here
we can see the voxel-wise encoding score for two Rank-1 models with FIR basis and 
different design matrices: separate design on the Y-axis and classical
design on the X-axis. Although both models give similar results, a Wilcoxon
signed-rank test on the leave-one-session-out cross-validation score confirmed the superiority of the separate designs model
in this dataset with p-value $<10^{-3}$.

\begin{figure}
\centering
\includegraphics[width=\linewidth]{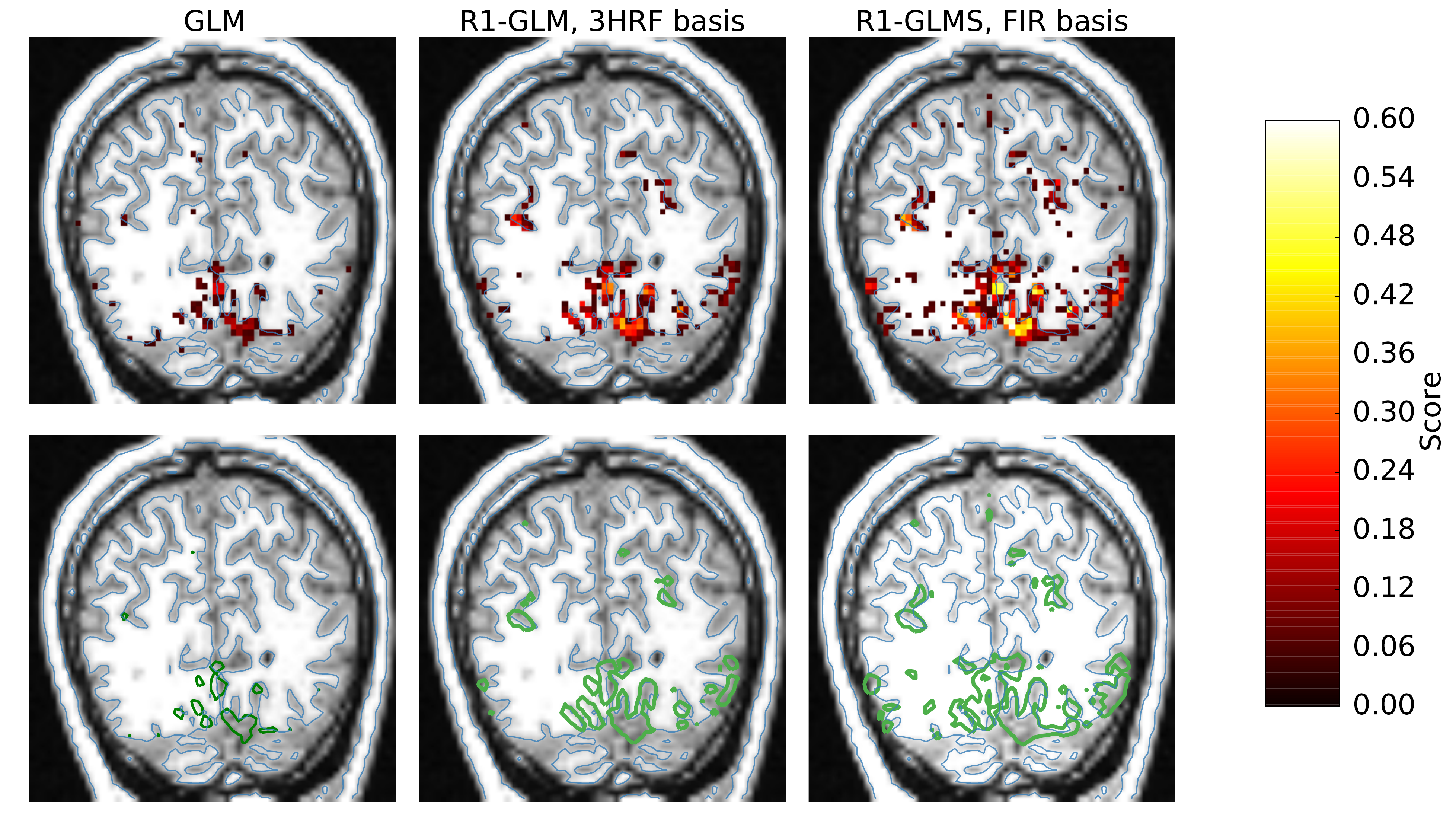}
\caption{\label{fig:brain_img}
Voxel-wise encoding scores on a single acquisition slice for different estimation methods (first dataset).
The metric is Pearson correlation.  
In the upper column, the voxel-wise score is thresholded at a value of 
0.045 (p-value $< 0.05$), 
while in the bottom row the 0.055 contour (p-value $< 0.001$) for the same data is shown as a green line. 
Despite lacking proper segmentations of visual areas,
the estimation methods produce results that highlight 
meaningful regions of interest around the calcarine fissure.
This is particularly visible in the third 
column where our method R1-GLMS produces results with higher
sensitivity
than the standard GLM method. In the bottom row it can be seen how the top performing voxels
follow well the folding of the gray matter.
}
\end{figure}

In Figure~\ref{fig:brain_img} we can see the voxel-wise encoding score on a single
acquisition slice. In the upper column, the score is plotted on each voxel and
thresholded at a value of 0.045, which would correspond to a p-value $< 0.05$
for testing non-correlation assuming each signal is normally distributed,
while in the bottom row the 0.055 contour (p-value $< 0.001$) for the same
data is shown as a green line. Here it can be seen how the top performing
voxels follow the gray matter. A possible hypothesis to explain the increase of the encoding score
between the method R1-GLMS with FIR basis and the same method with 3HRF basis
could be related either to the shape of the HRF deviating more from a canonical
shape in lateral visual areas or to the higher signal-to-noise ratio often found in the visual cortex when compared to lateral visual areas.

\subsection{Dataset 2: decoding of potential gain levels}

The mean decoding score was computed over 50 random splittings of the data,
with a test set of size 10\%. The decoding regression model consisted of
univariate feature selection (ANOVA) followed by a Ridge regression
classifier as implemented in scikit-learn~\cite{Pedregosa2011}. Both
parameters, number of voxels and amount of $\ell_2$ regularization in Ridge regression,
were chosen by cross-validation. 

\begin{figure}[tb]
\centering
\includegraphics[width=1.\linewidth]{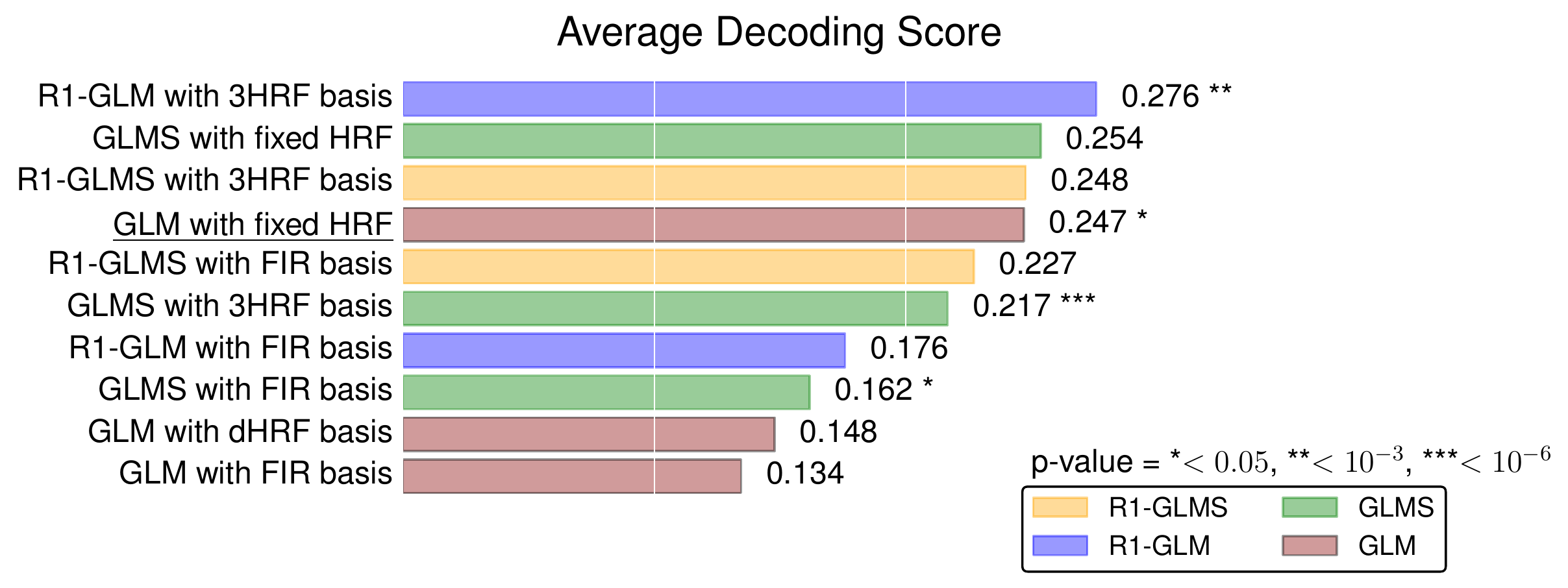}
\caption{\label{fig:decoding_scores}
Averaged decoding score for the different method considered 
(higher is better) on the second dataset. The metric is Kendall tau. Methods that perform constrained HRF estimation significantly outperform 
methods that use a fixed (reference) HRF. In particular,
the best performing method is the R1-GLM with 3HRF basis, followed by the R1-GLMS with 3HRF basis. 
In underlined typography is the GLM with a fixed HRF which is the method
used by default in most software distributions.
As in Figure~\ref{fig:encoding_scores}, a Wilcoxon signed-rank test is performed and the p-value reported between a given method and the next method in the ordered result list
to assess whether the difference in score is significant.
}
\end{figure}

The mean score for the 10 models considered can be seen in
Figure~\ref{fig:decoding_scores}. Similarly to how we assessed superiority of
a given method in encoding, we will say that a given method outperforms
another if the paired difference of both scores (this time across folds) is
significantly greater than zero. This is computed by performing a Wilcoxon
signed rank test across voxels. For this reason we report p-values
together with the mean score in Figure~\ref{fig:decoding_scores}.

As was the case in encoding, Rank-1 constrained methods obtain the highest
scores. In this case however, methods with 3HRF basis outperform methods using
FIR basis. This can be explained by factors such as smaller sample size of
each of the runs, smaller number of trials in the dataset and experimental design.

% An alternative approach, that has proven to increase robustness of the
% estimates, consists in performing a parcel-wise estimation of the
% HRF~\cite{Badillo2013}. While it would be possible to combine a rank-1
% constraint across conditions with a parcel model of the HRF, this lies
% outside the scope of this paper and is left for future work.

\section{Discussion}

We have compared different HRF modeling techniques and examined their
generalization score on two different datasets: one in which the main task was
an \emph{encoding} task and one in which it was a \emph{decoding} task. We
compared 10 different methods that share a common formulation within the
context of the General Linear Model. This includes models with canonical and
separate designs, with and without HRF estimation constrained by a basis set,
and with and without rank-1 constraint. We have focused on voxel-independent
models of the HRF, possibly constrained by a basis set, and have omitted for
efficiency reasons other possible models such as Bayesian
models~\cite{marrelec2003robust,Ciuciu2003,Makni2005} and regularized
methods~\cite{Goutte2000,Casanova2008}.

Other models such as spatial models~\cite{vincent2010spatially}, and 
\mbox{multi-subject}
methods~\cite{Zhang2012,Zhang2013} that adaptively learn the HRF
across several subjects are outside the scope of this work.  The latter models
are more relevant in the case of standard group studies and second level
analysis.

Our first dataset consists of an encoding study and
revealed that it is possible to boost the encoding score
by appropriately modeling the HRF. We used two different metrics to assess the quality of our estimates. The first metric is the fraction of correctly identified images by an encoding model. For this we computed the activation coefficients on both the training and validation dataset. We then learned a predictive model of the activation coefficients from the stimuli. This was used to identify a novel image from a set of 120 potential images from which the activation coefficients were previously computed. The benefits range from 0.9\% points to 8.2\% points across R1-constrained methods and subjects. The best-performing model in this task is the R1-GLM with FIR basis. The second metric is the Pearson correlation.
By considering the voxel-wise score on a
full brain volume we observed that the increase in performance obtained by
estimating the HRF was not homogeneous across voxels and more important for
voxels that already exhibited a good score with a classical design (GLM) and a
fixed HRF. The best-performing method is the Rank-1 with separate designs
(R1-GLMS) and FIR basis model, providing a significant improvement over the
second best-performing model. We also found substantial variability of
the shape in the estimated HRF within a single subject and a single task. 

% This important spatial variability of the estimated HRFs justifies a
% voxelwise estimation of the HRF without any spatial regularization. 

The second dataset consists of a decoding task and the results confirmed that
constrained (rank-1) estimation of the HRF also increased the decoding score
of a classifier. The metric here is Kendall tau. However, in this case the
best performing basis was no longer FIR basis consisting of ten elements but
the three elements 3HRF basis (HRF and derivatives) instead, which can be
explained by factors such as differences in acquisition parameters,
\mbox{signal-to-noise} ratio or by the regions involved in the task.

A higher performance increase was observed when considering the correlation score within the encoding model. This 
higher sensitivity to a correct (or incorrect) estimation of the HRF
can be explained by the fact that the
estimation of the HRF is used to generate the BOLD signal on the test set. The metric
is the correlation between the generated signal and the BOLD signal.
It is thus natural to expect that a correct estimation of the HRF has a higher
impact on the results. 

 In the decoding setup, activation coefficients 
(beta-map) are computed but the evaluation metric is the accuracy at predicting the
stimulus type. The validation metric used for decoding is less sensitive to the HRF estimation procedure than the correlation metric from the encoding study, although it
allowed us to observe a statistically significant improvement.

% The rank-constrained model described in this paper assumes that the HRF is the
% same across conditions. This can be considered a limitation. It however
% applies naturally to experimental paradigms whenever the different stimuli are
% similar. A possible extension to this model would be to consider a setting in
% which  the conditions are divided into different groups (for example visual
% and  auditory) and only the conditions within a single group are linked
% together with a rank 1 constraint. This would result in as many HRFs estimated
% as the number of groups.

% A possible extension to this work would consist in determining the optimal
% ammount of basis functions to use for a given dataset. Indeed, we have observed
% that smaller basis sets obtain a better score 

\section{Conclusion}

We have presented a method for the joint estimation of HRF and
activation coefficients within the GLM framework. Based on ideas from previous
literature~\cite{Makni2008,vincent2010spatially} we assume the HRF to be equal
across conditions but variable across voxels. Unlike previous work, we cast
our model as an optimization problem and propose an efficient
algorithm based on quasi-Newton methods. We also extend this approach to the setting of GLM
with separate designs.

We quantify the improvement in terms of generalization score in both encoding
and decoding settings. Our results show that the rank-1 constrained method
(R1-GLM and R1-GLMS) outperforms competing methods in both encoding and
decoding settings. 

\textbf{Acknowledgements}
This work was supported by grants IRMGroup ANR-10-BLAN-0126-02 and BrainPedia
ANR-10-JCJC 1408-01. We would like to thank our colleagues Ronald Phlypo and Gael
Varoquaux for fruitful discussions.

\bibliographystyle{elsarticle-harv}

\bibliography{biblio}

\end{document}